# High-Performance Cloud Computing: A View of Scientific Applications

Christian Vecchiola[1], Suraj Pandey[1], and Rajkumar Buyya[1,2]

[1]Cloud computing and Distributed Systems (CLOUDS) Laboratory
Department of Computer Science and Software Engineering
The University of Melbourne, Australia
{csve, spandey, raj}@csse.unimelb.edu.au

[2]Manjrasoft Pty Ltd., Melbourne, Australia

*Abstract*— Scientific computing often requires the availability of a massive number of computers for performing large scale experiments. Traditionally, these needs have been addressed by using high-performance computing solutions and installed facilities such as clusters and super computers, which are difficult to setup, maintain, and operate. Cloud computing provides scientists with a completely new model of utilizing the computing infrastructure. Compute resources, storage resources, as well as applications, can be dynamically provisioned (and integrated within the existing infrastructure) on a pay per use basis. These resources can be released when they are no more needed. Such services are often offered within the context of a Service Level Agreement (SLA), which ensure the desired Quality of Service (QoS). Aneka, an enterprise Cloud computing solution, harnesses the power of compute resources by relying on private and public Clouds and delivers to users the desired QoS. Its flexible and service based infrastructure supports multiple programming paradigms that make Aneka address a variety of different scenarios: from finance applications to computational science. As examples of scientific computing in the Cloud, we present a preliminary case study on using Aneka for the classification of gene expression data and the execution of fMRI brain imaging workflow.

## I. INTRODUCTION

Scientific computing involves the construction of mathematical models and numerical solution techniques to solve scientific, social scientific and engineering problems. These models often require a huge number of computing resources to perform large scale experiments or to cut down the computational complexity into a reasonable time frame. These needs have been initially addressed with dedicated high-performance computing (HPC) infrastructures such as clusters or with a pool of networked machines in the same department, managed by some "CPU cycle scavenger" software such as Condor [1]. With the advent of Grid computing [2] new opportunities became available to scientists: in a complete analogy with the power Grid [3], the computing Grid could offer on demand the horse power required to perform large experiments, by relying on a network of machines, potentially extended all over the world. Computing Grids introduced new capabilities such as dynamic discovery of services, the ability of relying on a larger number of resources belonging to different administrative domains and of finding the best set of machines meeting the requirements of applications. The use of Grids for scientific computing [4] has become so successful that many international projects led to the establishment of world-wide infrastructures available for computational science. The Open Science Grid [5], originally conceived for facilitating data analysis for the Large Hadron Collider, actually hosts 25000 machines and provides support for data intensive research for different disciplines such as biology, chemistry, particle physics, and geographic information systems. Enabling Grid for E-SciencE (EGEE) [6] is an initiative funded by the European Commission that connects more than 91 institutions in Europe, Asia, and United States of America, to construct the largest multi-science computing Grid infrastructure of the world. TeraGRID [7] is an NSF funded project that provides scientists with a large computing infrastructure built on top of resources at nine resource provider partner sites. It is used by 4000 users at over 200 universities that advance research in molecular bioscience, ocean science, earth science, mathematics, neuroscience, design and manufacturing, and other disciplines. These are only the most representative examples of scientific Grid computing.

Even though the widespread use of Grid technologies in scientific computing is demonstrated by the huge amount of projects served by the aforementioned computing Grids, some issues still make the access to this technology not as easy as depicted. Some issues are bureaucratic: being these Grids shared worldwide, research groups have to submit a proposal describing the type of research they want to carry out. This approach leads to a competitive use of scientific Grids, and minor research projects could not get access to them. Other issues are technical and more important: in most of the cases scientific Grids feature a pre-packaged environment in which applications will be executed, sometimes specific tools and APIs have to be used and there could be limitations on the hosting operating systems or on the services offered by the runtime environment. Even though Grid computing fosters the dynamic discovery of services and a wide variety of runtime environments for applications, in practice a limited set of options is available for scientists, and sometimes they could not be elastic enough to cover their needs. A practical example, involves the use of specific software that could not be available in the runtime environment were applications are executed. In general, applications that run on scientific Grids are implemented as bag of tasks applications, workflows, and MPI (Message Passing Interface) [8] parallel processes. Some scientific experiments could not fit into these models and have to be reorganized or redesigned to make use of scientific Grids. Whereas the bureaucratic issues can be a minor problem, the technical ones could constitute a fundamental obstacle for scientific computing. In this sense, the approach based on virtualized technologies proposed by the PlanetLab [9] could be of great help. PlanetLab is an open platform for developing, deploying, and accessing planetary-scale services. Users are given a slice that is virtual machine access on a set of nodes in the PlanetLab infrastructure. Hence, a slice can be fully customizable for the specific use. At present, PlanetLab is mostly used as a testbed for computer networking and distributed system research and it is only accessible to the infrastructure is granted only to persons affiliated with corporations and universities that host PlanetLab node. This makes its use for computational science quite limited.

Cloud computing [10][11], the current emerging trend in delivering IT services, can address many of the aforementioned problems. By means of virtualization technologies, Cloud computing offers to end users a variety of services covering the entire computing stack, from the hardware to the application level, by charging them on a pay per use basis. Another important feature, from which scientists can benefit, is the ability to scale up and down the computing infrastructure according to the application requirements and the budget of users. By using Cloud based technologies scientists can have easy access to large distributed infrastructures and completely customize their execution environment, thus providing the perfect setup for their experiments. Moreover, by renting the infrastructure on a pay per use basis, they can have immediate access to required resources without any capacity planning and they are free to release them when resources are no longer needed. Cloud computing provides a flexible mechanism for delivering IT services at each level of the computing stack: from the hardware level to the application level. Hardware appliances and applications are provisioned by means of hardware virtualization and software-as-a-service solutions, respectively. This makes the spectrum of options available to scientists wide enough to cover any specific need for their research.

The interest towards Cloud computing solutions is rapid growing. As a result, they have already been adopted in different scenarios such as social networking, business applications, and content delivery networks. At present, the use of Cloud computing in computational science is still limited, but the first steps towards this goal have been already done. This year the Department of Energy (DOE) National Laboratories started exploring the use of cloud services for scientific computing. On April 2009, Yahoo Inc. announced that it has extended its partnership with the major top universities in United States of America to advance Cloud computing research and applications to computational science and engineering. One of the first cloud-based infrastructures for computational science, Science Cloud [12], has been already deployed by the joint efforts of the University of Chicago, the University of Illinois, Purdue University, and Masaryk University. From a research point of view, initial studies have been conducted on the feasibility of using computing clouds for scientific computing. Some studies investigated the benefit of using Cloud computing technologies by analyzing the performances of HPC scientific applications [13] or the cost of performing scientific experiments [14] on the Amazon Cloud infrastructure.

Different solutions are available to move from the traditional science Grids and embrace the Cloud computing paradigm. Some vendors, such as Amazon Web Services and VMWare base their offering on hardware level virtualization and provide bare compute and storage resources on demand. Google AppEngine and Microsoft Azure are more focused on application level virtualization by enforcing a specific application model that leverage their large infrastructure and scale up and down on demand. Other solutions provide end users with a platform for developing Cloud computing applications that can rely on, or compose, some of the existing solutions thus providing a better Quality of Service to the end user. Aneka [15] is a Cloud computing platform for developing applications that can scale on demand by harnessing the CPU cycles of virtual resources, desktop PCs, and clusters. Its support for multiple programming models provides scientists with different options for expressing the logic of their applications: bag of tasks, distributed threads, dataflow, or MapReduce [16]. Its service oriented architecture provides users with an extremely customizable infrastructure that can meet the desired Quality of Service for applications.

The rest of the paper is organized as follows: first, we provide an overview of Cloud computing by defining the reference model and the key elements of this paradigm. Then, we will introduce Aneka and provide a detailed discussion of its features by highlighting how it can support computational science. As case studies, we will present the classification of gene expression data and the execution of scientific workflows on the Amazon EC2 public cloud. Final thoughts and key observations about the future directions of Cloud computing, as a valid support for scientific computing, are discussed at the end.

II. THE RISE OF THE CLOUDS

The term Cloud computing encompasses many aspects that range from the experience that end users have with the new opportunities offered by this technology to the implementation of systems that actually make these opportunities a reality. In this section, we will provide a characterization of what Cloud computing is, introduce a reference model for Cloud computing, and identify the key services that this new technology offers.

*A. Cloud Definition*

Although, the term Cloud computing is too broad to be captured into a single definition it is possible to identify some key elements that characterize this trend. Armbrust et al. [10] observe that *"Cloud computing refers to both the applications delivered as services over the Internet and the hardware and system software in the datacenters that provide those services"*. They then identify the Cloud with both the hardware and the software components of a datacenter. A more structured definition is given by Buyya et al. [17] who define a Cloud as a *"type of parallel and distributed system consisting of a collection of interconnected and virtualized computers that are dynamically provisioned and presented as one or more unified computing resources based on service-level agreement"*. One of the key features characterizing Cloud computing is the ability of delivering both infrastructure and software as services. More precisely, it is a technology aiming to deliver on demand IT resources on a pay per use basis. Previous trends were limited to a specific class of users, or specific kinds of IT resources. Cloud computing aims to be global: it provides the aforementioned services to the mass, ranging from the end user that hosts its personal documents on the Internet, to enterprises outsourcing their entire IT infrastructure to external datacenters.

### B. Cloud Computing Reference Model

Figure 1 gives an overview of the scenario envisioned by Cloud computing. It provides a layered view of the IT infrastructure, services, and applications that compose the Cloud computing stack. It is possible to distinguish four different layers that progressively shift the point of view from the system to the end user.

The lowest level of the stack is characterized by the physical resources on top of which the infrastructure is deployed. These resources can be of different nature: clusters, datacenters, and spare desktop machines. Infrastructures supporting commercial Cloud deployments are more likely to be constituted by datacenters hosting hundreds or thousands of machines, while private Clouds can provide a more heterogeneous environment, in which even the idle CPU cycles of spare desktop machines are used to leverage the compute workload. This level provides the "horse power" of the Cloud.

The physical infrastructure is managed by the core middleware layer whose objectives are to provide an appropriate runtime environment for applications and to exploit the physical resources at best. In order to provide advanced services, such as application isolation, quality of service, and sandboxing, the core middleware can rely on virtualization technologies. Among the different solutions for virtualization, hardware level virtualization and programming language level virtualization are the most popular. Hardware level virtualization guarantees complete isolation of applications and a fine partitioning of the physical resources, such as memory and CPU, by means of virtual machines. Programming level virtualization provides sandboxing and managed execution for applications developed with a specific technology or programming language (i.e. Java, .NET, and Python). On top of this, the core middleware provides a wide set of services that assist service providers in delivering a professional and commercial service to end users. These services include: negotiation of the quality of service, admission control, execution management and monitoring, accounting, and billing. Together with the physical infrastructure, the core middleware represents the platform on top of which the applications are deployed in the Cloud. It is very rare to have direct user level access to this layer. More commonly, the services delivered by the core middleware are accessed through a user level middleware. This provides environments and tools simplifying the development and the deployment of applications in the Cloud. They are: web 2.0 interfaces, command line tools, libraries, and programming languages. The user-level middleware constitutes the access point of applications to the Cloud.

### C. Cloud Computing Services Offering

The wide variety of services exposed by the Cloud computing stack can be classified and organized into three major offerings that are available to end users, scientific institutions, and enterprises. These are: Infrastructure as a Service (IaaS), Platform as a Service (PaaS), and Software as a Service (SaaS). Figure 2 provides such categorization.

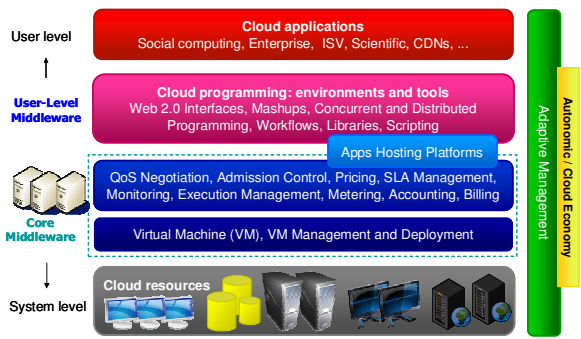

Figure 1. Cloud computing layered architecture.

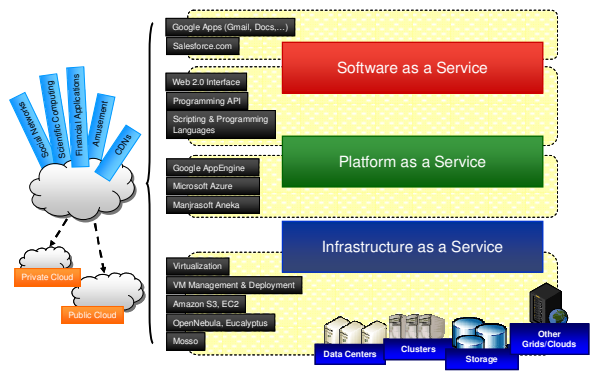

Figure 2. Cloud computing offerings by services.

Infrastructure as a Service or Hardware as a Service (HaaS) are terms that refer to the practice of delivering IT infrastructure based on virtual or physical resources as a commodity to customers. These resources meet the end user requirements in terms of memory, CPU type and power, storage, and, in most of the cases, operating system. Users are billed on a pay per use basis and have to set up their system on top of these resources that are hosted and managed in datacenters owned by the vendor. Amazon is one of the major players in providing IaaS solutions. Amazon Elastic Compute Cloud (EC2) provides a large computing infrastructure and a service based on hardware virtualization. By using Amazon Web Services, users can create Amazon Machine Images (AMIs) and save them as templates from which multiple instances can be run. It is possible to run either Windows or Linux virtual machines and the user is charged per hour for each of the instances running. Amazon also provides storage services with the Amazon Simple Storage Service (S3), users can use Amazon S3 to host large amount of data accessible from anywhere.

Platform as a Service solutions provide an application or development platform in which users can create their own application that will run on the Cloud. PaaS implementations provide users with an application framework and a set of API that can be used by developers to program or compose applications for the Cloud. In some cases, PaaS solutions are generally delivered as an integrated system offering both a development platform and an IT infrastructure on top of which applications will be executed. The two major players adopting this strategy are Google and Microsoft.

Google AppEngine is a platform for developing scalable web applications that will be run on top of server infrastructure of Google. It provides a set of APIs and an application model that allow developers to take advantage of additional services provided by Google such as Mail, Datastore, Memcache, and others. By following the provided application model, developers can create applications in Java, Python, and JRuby. These applications will be run within a sandbox and AppEngine will take care of automatically scaling when needed. Google provides a free but limited service while utilizes daily and per minute quotas to meter and price applications requiring a professional service. Azure is the solution provided by Microsoft for developing scalable applications for the Cloud. It is a cloud service operating system that serves as the development, run-time, and control environment for the Azure Services Platform. By using the Microsoft Azure SDK, developers can create services that leverage the .NET Framework. These services have to be uploaded through the Microsoft Azure portal in order to be executed on top of Windows Azure. Additional services, such as workflow execution and management, web services orchestration, and access to SQL data stores, are provided to build enterprise applications.

Aneka [15], commercialized by Manjrasoft, is a pure PaaS implementation and provides end users and developers with a platform for developing distributed applications for the Cloud by using the .NET technology. The core value of Aneka is a service oriented runtime environment – the Aneka container – that is deployed on both physical and virtual infrastructures and allows the execution of applications developed by means of different programming models. Aneka provides a Software Development Kit (SDK) helping developers to create cloud applications on any language supported by the .NET runtime and a set of tools for setting up and deploying clouds on Windows and Linux based systems. Being a pure PaaS solution, Aneka does not provide an IT hardware infrastructure to build computing Clouds, but system administrator can easily set up Aneka Clouds by deploying the Aneka containers on clusters, datacenters, simple desktop PCs, or even bundled within Amazon Machine Images.

Software as a Service solutions are at the top end of the Cloud computing stack and they provide end users with an integrated service comprising hardware, development platforms, and applications. Users are not allowed to customize the service but get access to a specific application hosted in the Cloud. Examples of the SaaS implementations are the services provided by Google for office automation, such as Google Document and Google Calendar, which are delivered for free to the Internet users and charged for professional quality services.

TABLE I. CLOUD COMPUTING SOLUTIONS FEATURE COMPARISON

| Properties | Amazon EC2 | Google AppEngine | Microsoft Azure | Manjrasoft Aneka |
|---|---|---|---|---|
| Service Type | IaaS | IaaS – PaaS | IaaS – PaaS | PaaS |
| Support for (value offer) | Compute/Storage | Compute(web applications) | Compute/Storage | Compute |
| Value Added Provider | Yes | Yes | Yes | Yes |
| User access Interface | Web APIs and Command Line Tools | Web APIs and Command Line Tools | Azure Web Portal | Web APIs, Custom GUI |
| Virtualization | OS on Xen Hypervisor | Application Container | Service Container | Service Container |
| Platform (OS & runtime) | Linux, Windows | Linux | .NET on Windows | .NET/Mono on Windows, Linux, MacOS X |
| Deployment Model | Customizable VM | Web apps (Python, Java, JRuby) | Azure Services | Applications (C#, C++, VB, ….) |
| If PaaS, ability to deploy on 3rd party IaaS | N.A. | No | No | Yes |

Examples of commercial solutions are Salesforce.com and Clarizen.com, which respectively provide on line CRM and project management services.

Table I gives a feature comparison of some of the most representative players in delivering IaaS/PaaS solution for Cloud computing. In the rest of the paper we will mostly concentrate on Aneka and how it can be used to support scientific computing in the Cloud.

## III. ANEKA

Aneka[1] is a software platform and a framework for developing distributed applications on the Cloud. It harnesses the computing resources of a heterogeneous network of desktop PCs and servers or datacenters on demand. Aneka provides developers with a rich set of APIs for transparently exploiting such resources and expressing the logic of applications by using a variety of programming abstractions. System administrators can leverage a collection of tools to monitor and control the deployed infrastructure. This can be a public cloud available to anyone through the Internet, or a private cloud constituted by a set of nodes with restricted access within an enterprise.

The flexible and service-oriented design of Aneka and its fully customizable architecture make Aneka Clouds able to support different scenarios. Aneka Clouds can provide the pure compute power required by legacy financial applications, can be a reference model for teaching distributed computing, or can constitute a more complex network of components able to support the needs of large scale scientific experiments. This is also accomplished by the variety of application programming patterns supported through an extensible set of programming models. These define the logic and the abstractions available to developers for expressing their distributed applications. As an example, in order to run scientific experiments it is possible to rely on a classic bag of tasks model, or to implement the application as a collection of interacting threads or MPI processes, a set of interrelated tasks defining a workflow, or a collection of MapReduce tasks. If the available options do not meet the requirements, it is possible to seamlessly extend the system with new programming abstractions.

Aneka Clouds can be built on top of different physical infrastructures and integrated with other Cloud computing solutions such as Amazon EC2 in order to extend on demand their capabilities. In this particular scenario, Aneka acts as a middleman mitigating the access to public clouds from user applications. It operates as an application service provider that, by using fine and sophisticated pricing policies, maximizes the utilization of the rented virtual resources and shares the costs among users. Of a particular importance are then, the accounting and pricing services and how they operate when Aneka integrates public clouds.

Figure 3 gives an architectural overview of Aneka. In order to develop cloud computing applications developers are provided with a framework that is composed by a software development kit for programming applications, a management kit for monitoring and managing Aneka Clouds and a configurable service based container that constitute the building blocks of Aneka Clouds. In this section we will mostly focus on three key features: the architecture of Aneka clouds, the application model, and the services available for integrating Aneka with public clouds.

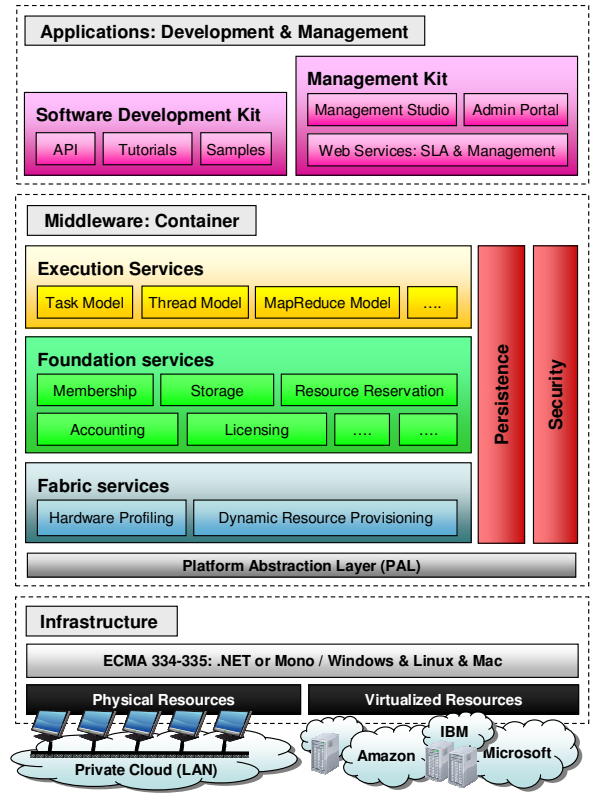

Figure 3. Aneka architecture.

### A. Aneka Clouds

The Aneka cloud is a collection of software daemons – called *containers* – that can be hosted on either physical or virtual resources and that are connected through a network such as the Internet or a private intranet. The Aneka container is the building block of the entire system and exposes a collection of services that customize the runtime environment available for applications.

It provides the basic management features for a single node and leverages the hosted services to perform all the other operations. We can identify fabric and foundation services. Fabric services directly interact with the node through the Platform Abstraction Layer (PAL) and perform hardware profiling and dynamic resource provisioning. Foundation services identify the core system of the Aneka infrastructure; they provide a set of basic features on top of which each of the Aneka containers can be specialized to perform a specific set of tasks. One of the key features of Aneka is the ability to provide multiple ways of expressing distributed applications by offering different programming models; execution services are mostly concerned with providing the middleware with an implementation for these models. Additional services such as persistence and security are transversal to the entire stack of services that are hosted by the container.

---
[1] Originally Aneka began as a third generation enterprise Grid initiative [28] in 2006 and its goals were in line with what is promised by the Cloud computing paradigm. It then rapidly emerged as a PaaS for Clouds.

The network of containers can be the result of different deployment scenarios: it can represent a private cloud completely composed by physical machines (desktop PCs and clusters) within the same administrative domain such as an enterprise or a university department. On the other hand, a totally virtual infrastructure is possible and the entire Aneka Cloud can be hosted on a public cloud such as Amazon EC2 or a private datacenter managed by Eucalyptus [18]. Hybrid systems are also allowed and they are the most frequent ones. In this case, the local infrastructure is extended with additional virtual resources as depicted in Figure 4.

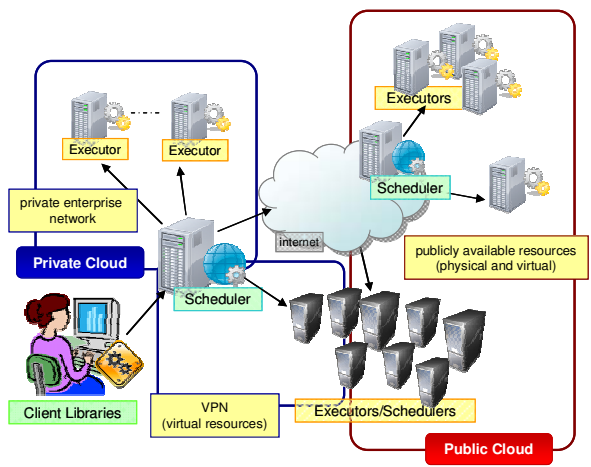

Figure 4.  Aneka deplyoment on hybrid clouds.

Aneka Clouds can scale on demand and provision additional nodes or release some of them when they are no longer needed. These nodes can either be virtual or physical resources. Physical nodes can be released by the network by simply shutting down the container hosted in the node, while in the case of the virtual resources it is also necessary to terminate the virtual machine hosting the container. This process can be performed manually or by the scheduler as a result of the elastic and autonomic management of the status of the Cloud. Except for the provisioning policy, there is no difference between containers hosted within a virtual machine or a physical resource since all the hardware related tasks are encapsulated in the Platform Abstraction Layer (PAL). As described in Figure 3, the provisioning module belongs to the Fabric Services and exposes its services to the other components that operate indifferently.

A set of services are always deployed within Aneka Clouds. Apart from the Fabric Services, the operative core of the container is represented by the Foundation Services, which perform the basic operations for managing the Aneka Cloud. Among these, the Membership Services play a key role in keeping track of all the nodes belonging to the Cloud and providing a registry that can be used for dynamically discovering the services available in the network or nodes with a specific configuration or operating system. For example, they can be used by scheduling services to locate all the nodes that can support the execution of a given programming model. Other components provide basic features such as support for file transfer and resource reservation for privileged execution.

The customization of the Aneka container takes place on top of the Foundation Services. Even though, the Aneka container is configurable and customizable at any layer, Execution Services are the one that more commonly differentiate the purpose of a node. A classic configuration, which is depicted in Figure 4, features a deployment where scheduling services are installed on a limited number of nodes and the majority of the containers are configured to be compute resources. This scenario identifies a master slave topology and it is only one of the possible options with Aneka, which could be suitable only for some programming models. A hierarchical topology based on schedulers and meta- schedulers can provide a better solution for large infrastructures and heavy load conditions.

This brief overview provides a general idea of the design principles that characterize the Aneka Cloud and its internal architecture. In the following, we will describe how these features are exploited by the Aneka application model for providing a customized runtime environment able to support different application programming patterns.

*B.  Aneka Application Model*

The Aneka Application Model defines the fundamental abstractions that constitute a distributed application hosted in the Aneka Cloud. It identifies the requirements that every specific implementation has to meet in order to be seamlessly integrated into Aneka and take advantage of all the available services hosted in the Cloud. The application model also specifies what the general requirements for the runtime environment that is expected to run applications that are built on top of a specific model.

Differently from other middleware implementations Aneka does not support single task execution, but any unit of user code is executed within the context of a distributed application. An application in Aneka is constituted by a collection of execution units whose nature depends on the specific programming model used. An application is the unit of deployment in Aneka and configuration and security operates at application level. Execution units constitute the logic of the applications. The way in which units are scheduled and executed is specific to the programming model they belong to. By using this generic model, the framework provides a set of services that work across all programming models supported: storage, persistence, file management, monitoring, accounting, and security.

In order to implement a specific programming model on top of Aneka developers have to:

- define the abstractions that will be used by software engineers to structure the distributed application and define its execution logic;
- provide the implementation of execution services that are required to manage the execution of the abstractions within the Aneka Cloud;
- implement a client component coordinating with the execution services that manages the execution from the client machine.

These are the components that are common to any different implementation of the programming model. The current release of Aneka supports four different programming models. These are: Task Model, Thread Model, MapReduce Model, and Parameter Sweep Model (PSM). Others, such as the Actor Model, the MPI Model, and Workflow are under development.

Table II, provides a feature comparison of these models and demonstrates the flexibility of the Aneka Application Model. For each of these models, the application type or scenario that naturally fit in that model are briefly described. The table provides a concise description of each model, which is represented in terms of application, execution units, and execution services within Aneka. It also provides a user and a system point of view.

*C. Accouting, Pricing, and Integration with Public Clouds*

Aneka provides an infrastructure that allows setting up private, public, and hybrid clouds. In a cloud environment, especially in the case of public and hybrid clouds, it is important to implement mechanisms for controlling resources and pricing their usage in order to charge users and maximize the utilization of the system while trying to minimize the costs. Accounting and pricing are the tasks that collectively implement a pricing mechanism for applications in Aneka.

The accounting service is responsible of keeping track of usage statistics of the systems and classifying them per user and application. This information is fundamental in order to estimate the cost that has to be charged to each user and to determinate how the applications are responsible of the user expenses. The current implementation of the accounting service is able to keep track of time spent by each execution unit of each application and to maintain the history of the execution of each unit. These data are then used by the selected pricing strategy to define the amount that has to be charged by the user. For example, a simple policy could be assigning a price to each resource and determinate the cost generated by each application by simply doing the weighted sum of all the execution units of the application. Other policies can take into account the specific services used by one application.

The role of these two components becomes even more important when Aneka Clouds are completely deployed or integrate with public clouds. In this case the use of virtual public resources incurs costs that have to be taken into account while determining the bill for users. Hybrid clouds constitute a challenging scenario: here, virtual resources are provisioned in order to meet the Service Level Agreement (SLA) signed with users. In order to face this challenge, Aneka provides an object model allowing third parties to seamlessly integrate different scheduling algorithms that can coordinate their activity with the resource provisioning service. The current implementation is still at an early stage and devises a model where the scheduler can access multiple resource pools keeping track in real time of the cost currently spent for each active instance. The basic strategy of the pool is trying to reuse as much as possible the instances already active in order to minimize the costs of the public virtual resources. Different scheduling algorithms can be plugged into this model; therefore, developers can provide multiple policies for deciding when to grow or shrink the set of nodes that constitute the Aneka Cloud.

TABLE II.  PROGRAMMING MODELS FEATURE COMPARISON

| Name | Scenario | Applications | Execution Units | Execution Services |
|---|---|---|---|---|
| Task Model | Independent bag of tasks applications | A collection of independent tasks. | Task interface, execute method. | Task scheduling service and task execution service. |
| Thread Model | Multithreaded applications | A collection of threads executed concurrently. | Any instance, any method. | Thread scheduling service and thread execution service |
| MapReduce Model | Data intensive applications | A map and a reduce functions and a large collection of data. | Map and Reduce Tasks | MapReduce scheduling and execution services, MapReduce storage service |
| PSM | Parameter sweeping applications | Task template with a collection of parameters. | Task template instance with a given set of parameter values | Built on top of the Task Model, no additional requirements |
| Workflow | Workflow applications | A collection of interrelated tasks composing a DAG. | Task instance | Built on top of the Task Model with additional requirements |
| MPI | Message passing applications | A collection of MPI processes that exchange messages. | MPI processes | MPI scheduling service, MPI execution service |
| Actors | Distributed active objects / agents | A collection of Actor instances interacting each other via message passing | Actor instances | Actor scheduling service, Naming service, Actor execution service hosting the actor theater |

## IV. CASE STUDIES

In this section, we will discuss two practical applications of scientific computing in the Cloud. Both the case studies have been implemented on top of the Amazon EC2 infrastructure. The first case study features the classification of gene expression datasets by using an Aneka Cloud while the second case presents the execution of an fMRI brain imaging workflow and compares its performance with the same experiment carried out on traditional Grids. In both of the two cases, a cost analysis on the usage of the Cloud is presented.

### A. Classification of Gene Expression Data

Gene expression profiling is then measurement of the activity – the expression – of thousands of genes at once, to create a global picture of cellular function. The analysis of profiles, which is the measurement of the activity of genes, helps researchers to identify the relationships between genes and diseases and how cells react to a particular treatment. One of the most promising techniques supporting the analysis of gene profiles is the DNA microarray technology [19], which can be particularly helpful in cancer prediction. One of the disadvantages of this technique is the huge amount of data produced: the DNA profile for each patient where thousands of genes are organized as an array and whose state (active or not) is indicated by specific color or a black spot in the array. For these reasons, the classification of these profiles for cancer diagnosis cannot be performed without the aid of computerized techniques.

Among the different classification methods, the CoXCS classifier [20] has demonstrated to be particularly effective in classifying gene expression data sets. CoXCS is a co-evolutionary learning classifier based on feature space partitioning. It extends the XCS model [21] by introducing a co-evolutionary approach. Figure 5 provides a schematic example of the internal logic of CoXCS: a collection of independent populations of classifiers are trained by using different partitions of the feature space within the training datasets. After a fixed number of iterations, selected classifiers from each of the independent populations are transferred to a different population according to some migration strategy. The evolution process is then repeated until a specific threshold is reached.

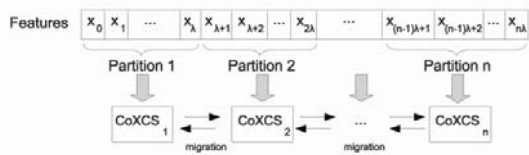

Figure 5. CoXCS architecture.

The internal architecture of CoXCS, based on feature space partitioning, not only outperforms the original XCS in classifying gene expression datasets, but also all the classic methods. Table III shows a performance comparison of different classification methods when applied to two sample gene expression datasets. As it can be noticed by the results obtained for the test phase the accuracy of CoXCS is definitely better than the ones achieved by other classifiers.

TABLE III. AREA UNDER COVER BASED ACCURACY COMPARISON

| Classifier | Mode | BRCA | Prostate |
|---|---|---|---|
| J48 | Train | 0.92 ± 0.06 | 1.00 |
| | Test | 0.35 ± 0.01 | 0.60 ± 0.10 |
| NBTree | Train | 1.00 | 1.00 |
| | Test | 0.65 ± 0.12 | 0.46 ± 0.04 |
| Random Forest | Train | 1.00 | 1.00 |
| | Test | 0.51 ± 0.01 | 0.60 ± 0.09 |
| Logistic Regression | Train | 1.00 | 0.50 |
| | Test | 0.85 ± 0.17 | 0.50 |
| Naïve Bayes Classifier | Train | 0.99 ± 0.01 | 1.00 |
| | Test | 0.90 ± 0.05 | 0.35 ± 0.04 |
| SVM | Train | 1.00 | 1.00 |
| | Test | 0.53 ± 0.04 | 0.51 ± 0.07 |
| XCS | Train | 0.50 | 0.50 |
| | Test | 0.50 | 0.50 |
| CoXCS | Train | 1.00 | 1.00 |
| | Test | **0.98 ± 0.02** | **0.70 ± 0.02** |

The only disadvantage of using CoXCS is the long computation time required to evolve the classifier into a stable form. The intrinsic parallelism of CoXCS allows for a distributed, and faster, implementation. Cloud-CoXCS is a Cloud-based implementation of CoXCS that leverage the Aneka Computing cloud to distribute the evolution of the independent populations of classifiers at each of the iterations. In order to quickly have a working prototype of Cloud-CoXCS, we implemented it as a strategy for the Offspring toolkit [22]. Offspring is a software environment allowing the rapid prototyping of *strategies*. These are client-based workflows that can be executed over Aneka and other middleware implementations.

The algorithm implemented in Cloud-CoXCS is the same as CoXCS; therefore, we expect to have the same accuracy and to obtain an almost linear speedup in the execution of the training sessions. In order to validate out assumptions, we deployed an Aneka Cloud on top of the Amazon EC2 infrastructure and performed some preliminary tests. These experiments allowed us to investigate the impact of the different setups on the execution time and the performance of Cloud-CoXCS. We performed two experiments: one varies the gene expression datasets; the other the type of instance of compute nodes deployed in Amazon EC2.

We did not change the parameter of CoXCS that have been fixed to the following values: 5000 individuals for each independent population; exploration/exploitation rate set to 0.3; 20 partitions for all datasets; migration rate set to 10% of the population size; 5 separate migration stages with 100 independent iterations for the evolutions of populations between migration stages.

Regarding the deployed infrastructure, two different Amazon images have been used: a master image and a slave image. The master image features an instance of the Aneka container hosting the scheduling and file staging services for the Task Model on a Windows Server 2003 operating system. The slave image hosts a container configured with the execution services deployed on a Red Hat Linux 4.1.2 (kernel: 2.6.1.7). The Cloud deployed for the experiments is composed by one master node and multiple slave nodes that have been added to the cloud on demand. Experiments have been done to compare different cloud setups. Two different image types have been tested for slave instances: *m1.small* and *c1.medium*. For the master node we used the *m1.small* instance type.

TABLE IV. ANEKA EC2 CLOUD SETUPS

| Feature | m1.small | c1.medium |
| --- | --- | --- |
| Cores | 1 | 2 |
| EC2 Computing Units | 2.5 | 5 |
| Memory | 1.7 GB | 1.7 GB |
| Slave Instances | 10 | 20 |
| Cost/Hour | 0.10 USD | 0.20 USD |

Table IV describes the characteristics of the two different cloud setups used for the experiments. It can be noticed that *c1.medium* instances are modeled as dual core machines and provide a computational power that is double compared to the one provided by *m1.small*. The computing power is expressed in EC2 Compute Units[2]. In both cases a complete parallelism at each stage is obtained because Aneka scheduler dispatches one task per core. Hence, *c1.medium* instances will receive two tasks to process each time.

In order to validate the experiments, we used the cross-validation technique. To support cross validation, the BRCA dataset has been configured with two folds, while the Prostate dataset has been divided into four folds. Table V reports the execution times recorded for the two setups.

TABLE V. EXECUTION TIMES (MINUTES)

| Setup | BRCA | | | | Prostate | |
| --- | --- | --- | --- | --- | --- | --- |
| Fold | 1 | 2 | 3 | 4 | 1 | 2 |
| M1.small | 08:26 | 10:00 | 10:00 | 9:00 | 35:13 | 40:44 |
| c1.medium | 10:42 | 10:04 | 15:17 | 11:42 | 52:48 | 53:48 |
| Delta | 2:16 | 0:04 | 05:17 | 02:42 | 17:35 | 13:04 |

The experiments performed on the two different setups show that in both of the two cases the execution times recorded for the Prostate dataset are approximately four times longer that the execution times recorded for the BRCA dataset. This is quite reasonable, since the number of genes BRCA samples is four times smaller than the number of samples in the Prostate dataset.

Another interesting aspect is the fact that execution with the *c1.medium* setup requires a major amount of time to complete. On average, about 32% more of the corresponding runs with the *m1.small* setup. Since the single CoXCS task that is executed is designed as a single threaded process, it does not take benefit from having the two cores. Moreover, the increase in execution time of the entire learning process can be explained by the fact that both tasks are competing for the shared resources such as cache, bus, memory, and file system. Another issue that has to be taken into account is the fact that CoXCS tasks are executed within the runtime environment provided by Aneka that is shared even on a dual core machine. For this reason, some small operations at the very basic level of the infrastructure are performed sequentially. Finally, CoXCS tasks are compute tasks that use a considerable amount of memory to perform the learning phase. This could create a major number of cache misses in a multi-core setup.

More interesting are the considerations about the budget spent for performing the experiments. Given the large number of partitions used, the duration of the single CoXCS task is quite limited. This allows completing the exploration of all the folds in less than one hour, in case of the BRCA dataset, and less than two hours in case of the Prostate dataset. These executions, even though with different timings, incur the same cost for both of the two setups. If we take into account, that there is a significant difference in the execution time (32% on average), the granularity offered by Amazon could not be enough to provide an efficient pricing model. The current accounting system implemented in Aneka keeps track of the execution of tasks by minute and it is designed to share the virtual resources among multiple users. By letting Aneka act as a middleman between the Cloud provider and the end-users, a more efficient billing strategy can be implemented.

*B. Functional Magnetic Resonance Imaging Workflows*

Brain imaging technologies focus on processing image data obtained from MRI (Magnetic Resonance Imaging) scanners. The processed images can be further analyzed by medical personnel and scientists. In particular, Functional Magnetic Resonance Imaging (fMRI) attempts to determine which parts of the brain reacts in response to some given stimulus. In order to achieve this goal, first, the images of the brain collected from MRI scanners have to be transformed in order to reduce anatomic variability that naturally differentiate one subject from another one. This process is known as *special normalization* or Image Registration (IR). Once this step is completed, the images are compared with the *atlas*, which is a reference image obtained as the average of all the subjects brain images, and as a final step, the specific fMRI analysis procedures are carried out.

Figure 6 provides a visual representation of the sequence of steps involved in the fMRI analysis. Of the whole fMRI process, only the spatial normalization, which involves a sequence of complex operations, has been modeled as a workflow. Such workflow, described in Figure 6, is both data and compute intensive in nature.

---

[2] An EC2 Compute Unit is a virtual metric used to express the computational power of an instance. One EC2 Compute Unit (ECU) provides the equivalent CPU capacity of 1.0-1.2 GHz 2007 Opteron or 2007 Xeon processor.

A typical scenario involves the use of 10 to 40 brain images for analysis, which are repeatedly carried out over different groups of subjects [25]. Each input image is around 16MB in size. For 20 images, the total input to the workflow is 640MB. The output data size of each process in the workflow ranges from 20MB to 40MB. In case of 40 brain images, the total size of data processed exceeds 20GB. The ideal execution time for a 1-subject IR workflow (done for estimating execution time for >1 subjects) is about 69 minutes on a single machine where there is no time spent for transferring data. In a distributed execution settings, where data transfer times and management overheads are non-trivial, the total time taken for execution increases significantly.

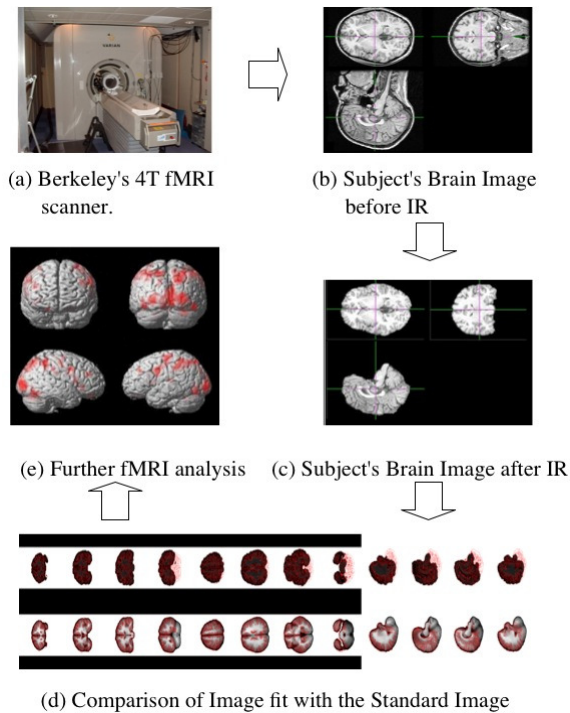

Figure 6. fMRI analysis.

The original experiment, featuring 20 subjects has been performed during the Second IEEE International Scalable Computing Challenge (SCALE 2009) held at CCGrid 2009 conference in Shanghai, China. The experiment results presented here are a part of the demonstration, which was one of the two winners of the competition.

The system deployed to run the experiments was completely hosted within the Amazon Cloud infrastructure. The execution of the workflow has been managed by the Gridbus Workflow Engine [26] that handled the execution of tasks in the workflow depicted in Figure 7. The experiment has been repeated with 2, 10 and 20 subjects and executed on the Amazon Cloud by using EC2 as a provider of computing resources and S3 for the storage of input data. The results of the execution have been compared with the execution of the same workflow in Grid'5000 [27], in which each compute node in the network served as both storage and compute resource. The metrics used to compare the results of the two executions is the *makespan* (difference between the submission time of the first submitted task and the output arrival time of the last exit task to be executed on the system) and execution cost of the workflow. The execution cost in Grid'5000 is assumed to be zero.

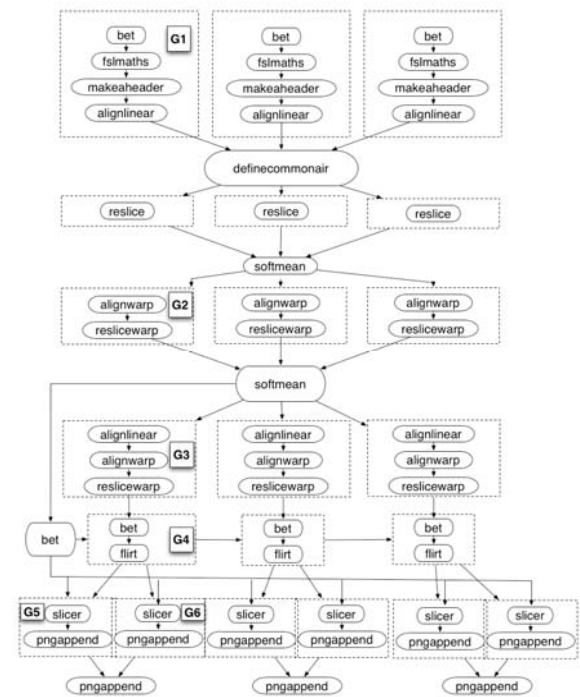

Figure 7. IR workflow structure.

Figure 8 compares the makespan of the workflow when the number of subjects used is varied. We observe that for large number of subjects, the makespan decreases when using EC2. For 2 subjects, the change in the makespan is not significant. This difference in makespan is mainly due to the shortening of the data-transfer time between the virtual nodes in EC2 as compared to the transfer between multiple physical sites in Grid'5000. For a large workflow (20 subjects) individual file transfer time gets cumulated, resulting in a significant difference in total makespan when compared to the results of Grid'5000.

Figure 9 compares the change in makespan versus the EC2 usage cost. The data transfer and the storage cost during execution were very minimal as the compute nodes were part of a Cloud datacenter. As the number of execution nodes is increased from 2 to 20 EC2 nodes, the makespan decreases significantly from 391 minutes to 107 minutes for a workflow analyzing 20 subjects. The cost of usage of Cloud nodes rose from $5.2 to $14.28. However, the ratio between the cost and the number of EC2 nodes used shows that: the total cost of computation of a large workflow (20 subjects) using 20 EC2 nodes would be $0.714/machine, as opposed to $2.6 when executing the same workflow using only 2 EC2 nodes. The average cost of usage per machine decreases as the number of resources provisioned increases from 2 to 20. Consequently, the overall application execution cost increased by not more than three times with a decrease in execution time by similar factor.

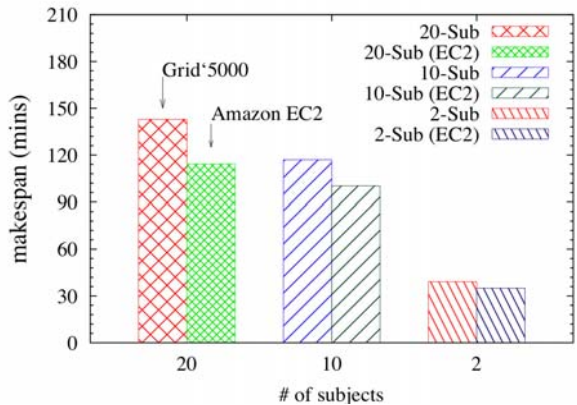

Figure 8. Makespan comparison between EC2 and Grid'5000 setups.

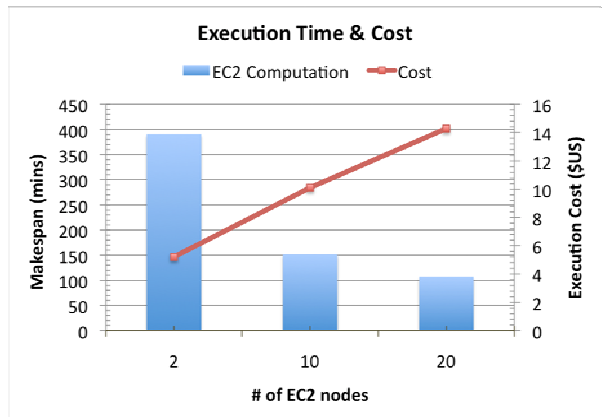

Figure 9. Makespan comparison between EC2 and Grid'5000 setups.

From our experiments, we conclude that large high performance applications can benefit from on-demand access and scalability of compute and storage resources provided by public Clouds. Hence, the increase in cost is subdued by the significant reduction in application execution time by making use of abundance of Cloud resources, which can be provisioned on demand.

## V. OBSERVATIONS AND THOUGHTS

The Cloud computing model introduces several benefits for end users, enterprises, service providers, and scientific institutions. The advantage of dynamically scaling the IT infrastructure on a pay per use basis and according to the real needs of applications, definitely constitute one of the major gains brought by Cloud computing. Moreover, by moving the IT infrastructure into the Cloud it is not necessary: (a) to bear costs derived from capacity planning for peak loads; (b) to statically acquire infrastructure due to the sporadic need of large computation power; and (c) to incur expensive administrative and maintenance costs. These issues are likely to be more important for enterprises and service providers that can maximize their revenue and cut costs. For what concerns end users, the most interesting aspect of Cloud computing resides in taking advantage of the multitude of applications already available and having their personal data and documents accessible from anywhere at anytime. On the other hand, scientific institutions can be more interested in PaaS and IaaS offerings that allow having complete control over the infrastructure used for scientific research and finely customizing their software systems according to the specific needs of the experiments to be performed. Cloud computing also ensures the desired Quality of Service, which is established by means of Service Level Agreements. This aspect can constitute an additional value, which could make scientists prefer computing Clouds to traditional Grids for experiments with additional constraints. For example, different types of analysis can be performed at different costs by optimizing the trade-off between the budget and the expected accuracy of results.

Due to its specific nature, Cloud computing introduces new challenges and new problems yet to be faced, especially from a legal and a security point of view. In the case of public clouds, systems, applications, and even personal data are hosted into datacenters owned by third parties. These datacenters are often placed into the more convenient geographic location for reducing maintenance and consumption costs. Such places could even be in a different country where different laws on the digital content apply. The same application can then be considered legal or illegal according to where it is hosted. In addition, privacy and confidentiality of data depends on the location of its storage. For example, confidentiality of accounts in a bank located in Switzerland may not be guaranteed by the use of datacenter located in United States. In order to address this issue some Cloud computing vendors have included the geographic location of the hosting as a parameter of the service level agreement made with the customer. For example, Amazon EC2 provides the concept of availability zones that identify the location of the datacenters where applications are hosted. Users can have access to different availability zones and decide where to host their applications. Security is another important issue, at the moment it is not clear which kind of measures, apart from the standard security tools, are taken to guarantee the privacy of data. While this issue is more compelling for enterprises and end users, there could be relevant implications even in the case of scientific computing: many scientific projects are often funded by federal bodies or directly by the government that often puts severe restrictions in the use and the management of sensible data.

## VI. Conclusions and Future Directions

We have discussed the potential opportunities and the current state-of-the-art of high-performance scientific computing on public clouds. The adoption of Cloud computing as a technology and a paradigm for the new era of computing has definitely become popular and appealing within the enterprise and service providers. It has also widely spread among end users, which more and more host their personal data to the cloud. For what concerns scientific computing, this trend is still at an early stage.

Science computing Grids such as Open Science Grid and EGEE already provide a large scale infrastructure, a set of well established methods and tools, and huge community of users. What could make interesting the use of computing Clouds for scientific institutions is the possibility of having a fully customizable runtime environment where they can carry out experiments. Other interesting opportunities arise, in considering the different available options in terms of Quality of Service for a given experiment. New and interesting scenarios can be explored, where scientists can decide the level of accuracy of their experiment or the specific partition of data to analyze according to the Service Level Agreement established with the Cloud provider. At present, some preliminary works have investigated the cost of doing science in the Cloud, by taking the Amazon EC2 and S3 infrastructure as case study. From an operational point of view, the first science computing cloud has been has been already deployed as a result of the joint efforts of a consortium of universities. The active interest of government bodies such as the Department of Energy in Cloud computing, will probably open pathways to the establishment of more science Clouds. A stronger adoption of Cloud computing for computational science will also contribute to advance research in other functional aspects such as security and jurisdiction. Many scientific projects are funded by the government bodies that sometimes impose significant restrictions on the use of data.

We also demonstrated some practical examples of doing science in the Cloud and presented the advanced features that Aneka provides for leveraging public and private Clouds to scale on demand according to the requirements of applications. Two case studies have been presented: the classification of gene expression data by using Aneka Cloud deployed on the Amazon EC2 infrastructure and the execution of scientific workflow on EC2. The preliminary considerations about the experiments performed show that the effective use of Cloud resources is really important and a trade-offs between cost and performance have to be carefully evaluated. This is where a platform like Aneka comes into picture. As a middleman, it gives access to cloud resources; maximizes their global usage; and provides end-users with a better pricing model. More detailed studies in this direction will definitely constitute the next step from this work.

### Acknowledgments

The authors would like to thank our colleagues Michael Kirley and Mani Abedini for their support and help in designing and implementing the Cloud enabled version of CoXCS, for the classification of gene expression data on top of the Amazon EC2 infrastructure. They have been very helpful in sketching the overall design of the classifier and precious in analyzing the results. We thank Xingchen Chu, Dileban Karunamoorthy, and Michael Mattess for their contribution to the recent development in Aneka; James E. Dobson from Dartmouth College, USA for providing us the fMRI application, Willian Voorsluys and Mustafizur Rahman, for their contribution in the development of the components of the workflow engine. Finally we would also like to express out gratitude to GRID'5000 project team for providing access to their resources.

### References

[1] D. Thain, T. Tannenbaum, and M. Livny, "Distributed computing in practice: The condor experience," Concurrency and Computation: Practice and Experience, vol. 17, pp. 323–356, February, 2005.

[2] Foster and C. Kesselman, The Grid: blueprint for a new computing infrastructure, I.Foster and C. Kesselman, Eds. San Francisco, CA, USA: Morgan Kaufmann Publishers Inc., November 1998.

[3] M. Chetty and R. Buyya, "Weaving Computational Grids: How Analogous Are They with Electrical Grids?," Computing in Science and Engineering (CiSE), vol. 4, July-August 2002, pp. 61–71, doi:10.1109/MCISE.2002.1014981.

[4] M. J. Chin, S. Harvey, S. Jha, and P. V. Coveney, "Scientific Grid Computing: The First Generation," Computing in Science and Engineering, vol. 7, 2005, . pp. 24–32.

[5] R. Pordes, D. Petravick, B. Kramer, D. Olson, M. Livny, A. Roy, P. Avery, K. Blackburn, T. Wenaus, F. Wurthwein, I. Foster, R. Gardner, M. Wilde, A. Blatecky, J. McGee, and R. Quick, "The open science Grid," Journal of Physics: Conference Series, vol. 78, no.1, 2007, pp. 012–057.

[6] F. Gagliardi, M.E. Begin, "EGEE - Providing a Production Quality Grid for e-science," Local to Global Data Interoperability - Challenges and Technologies, Sardinia, Italy, June, 2005

[7] C. E. Catlett, "TeraGrid: A Foundation for US Cyberinfrastructure," in Network and Parallel Computing, LCNS vol. 3779, H. Jin, D. Reed, and W. Jiang Eds., Springer Berlin / Heidelberg, 2005.

[8] W. Gropp, E. Lusk, and A.Skjellum, Using MPI: Portable Parallel Programming with the Message-passing Interface, ser. Scientific And Engineering Computation.MIT Press, 1994.

[9] T. Roscoe, "The PlanetLab platform", Peer-to-Peer Systems and Applications, LCNS vol. 3485, R. Steinmetz Ed., Springer Berlin / Heidelberg, 2005, pp. 567–581.

[10] M. Armbrust, A. Fox, R. Griffith, A. Joseph, R. Katz, A. Konwinski, G. Lee, D. Patterson, A. Rabkin, I. Stoica, M. Zaharia. Above the Clouds: A Berkeley View of Cloud computing. Technical Report No. UCB/EECS-2009-28, University of California at Berkley, USA, Feb. 10, 2009.

[11] R. Buyya, C. S. Yeo, S. Venugopal, J. Broberg, and I. Brandic. "Cloud Computing and Emerging IT Platforms: Vision, Hype, and Reality for Delivering Computing as the 5th Utility," Future Generation Computer Systems, vol. 25, no. 6, June 2009, pp 599–616, Elsevier Science, Amsterdam, The Netherlands.

[12] K. Keahey, T. Freeman. "Science Clouds: Early Experiences in Cloud computing for Scientific Applications," Cloud Computing and Its Applications 2008 (CCA-08), Chicago, IL. October 2008.

[13] C. Evangelinos, C. N. Hill, "Cloud Computing for Parallel Scientific HPC Applications: Feasibility of Running Coupled Atmosphere-Ocean Climate Models on Amazon's EC2," Cloud Computing and Its Applications 2008 (CCA-08), Chicago, IL. October 2008.

[14] E. Deelman, G. Singh, M. Livny, B. Berriman, and J. Good, "The cost of doing science on the cloud: the montage example," Proc. of the 2008 ACM/IEEE conference on Supercomputing (SC'08). Piscataway, NJ, USA: IEEE Press, 2008, pp. 1–12.

[15] C. Vecchiola, X. Chu, R. Buyya, "Aneka: a software platform for .NET-based Cloud computing," in High Performance & Large Scale Computing, Advances in Parallel Computing, W. Gentzsch, L. Grandinetti, G. Joubert Eds., IOS Press, 2009.

[16] J. Dean and S. Ghemawat, "Mapreduce: Simplified Data Processing on Large Clusters," Communications of ACM, vol. 51, no. 1, January 2008, pp. 107–113.

[17] R. Buyya, C.S. Yeo, and S. Venugopal, Market-Oriented Cloud Computing: Vision, Hype, and Reality for Delivering IT Services as Computing Utilities, Keynote Paper, in Proc. 10th IEEE International Conference on High Performance Computing and Communications (HPCC 2008), IEEE CS Press, Sept. 25–27, 2008, Dalian, China.

[18] D.Nurmi, R.Wolski, C.Grzegorczyk, G.Obertelli, S. Soman, L.Youseff, and D.Zagorodnov, "The Eucalyptus Open-source Cloud Computing System," Proc. 9th IEEE/ACM International Symposium on Cluster Computing and the Grid (CCGrid 2009), Shanghai, China: May 2009, pp. 124–131.

[19] G. Bucca, G. Carruba, A. Saetta, P. Muti, L. Castagnetta, and C.P. Smith, "Gene Expression Profiling of Human Cancer," Annals of New York Academy of Sciences, vol 1028, New York Academy of Sciences, United States, December, 2004.

[20] M. Abedini and M. Kirley, "CoXCS: A Coevolutionary Learning Classifier Based on Feature Space Partitioning," Proc. The 22nd Australasian Joint Conference on Artificial Intelligence (AI'09), Melbourne, Australia, December 1-4, 2009.

[21] S. Wilson, "Classifier Fitness Based on Accuracy," Evolutionary Computation, vol. 3. no. 2, 1995, 149–175.

[22] C. Vecchiola, M. Kirley, and R. Buyya, "Multi-Objective problem solving with Offspring on Enterprise Clouds," Proc. 10th International Conference on High Performance Computing in Asia Pacific Region (HPC Asia'09), Kaoshiung, Taiwan, March, 2009.

[23] I. Hedenfalk, D. Duggan, Y. Chen, M. Radmacher, M. Bittner, R. Simon, P. Meltzer, B. Gusterson, M. Esteller, O. P. Kallioniemi, B. Wilfond, A. Borg, and J. Trent, "Gene Expression profiles in hereditary breast cancer," N Engl J Med, vol. 344, no. 8, February 2001, pp. :539–548.

[24] M. M. Hossain, M. R. Hassan, and J. Bailey, "ROC-tree: A Novel Decision Tree Induction Algorithm Based on Receiver Operating Characteristics to Classify Gene Expression Data," Proc. SIAM International Conference on Data Mining, Atlanta, Georgia, USA, April 2008, pp. 455–465.

[25] S. Pandey, W. Voorsluys, M. Rahman, R. Buyya, J. Dobson, and K. Chiu, "A Grid Workflow Environment for Brain Imaging Analysis on Distributed Systems," Concurrency and Computation: Practice and Experience, Jul. 2009, doi:10.1002/cpe.1461.

[26] J. Yu and R. Buyya, "Gridbus workflow enactment engine," in Grid Computing: Infrastructure, Service, and Applications, L. Wang, W. Jie, and J. Chen Eds, CRC Press, Boca Raton, FL, USA, April 2009, pp. 119–146.

[27] F. Cappello and H. Bal, "Toward an international computer science Grid," Proc. 7th IEEE International Symposium on Cluster Computing and the Grid (CCGRID'07), pp 3–12, Rio, Brazil, 2007. IEEE.

[28] X. Chu, K. Nadiminti, C. Jin, S. Venugopal, R. Buyya, "Aneka: Next-Generation Enterprise Grid Platform for e-Science and e-Business Applications," Proc. 3rd IEEE International Conference on e-Science and Grid Computing, Bangalore, India, December, 2007.